\documentclass[12pt,onecolumn,draftclsnofoot]{IEEEtran}
\usepackage{amsfonts}
\IEEEoverridecommandlockouts
\ifCLASSINFOpdf
\else
\fi
\usepackage{multirow}
\usepackage{booktabs}
\usepackage{makecell}
\usepackage{graphicx}
\usepackage{subfigure}
\usepackage{cite}
\usepackage{color}
\usepackage[pagebackref=false,breaklinks=false,letterpaper=true,colorlinks,bookmarks=false]{hyperref}
\usepackage{amsmath}
\usepackage{amssymb}
\usepackage{array}
\usepackage{bm}
\usepackage{algorithm}
\usepackage{algorithmic}
\usepackage{setspace}

\usepackage{listings}
\usepackage{color}

\definecolor{dkgreen}{rgb}{0,0.6,0}
\definecolor{gray}{rgb}{0.5,0.5,0.5}
\definecolor{mauve}{rgb}{0.58,0,0.82}

\usepackage{cite}
\usepackage{lettrine} 

\title{Automatic Modulation Classification Using Involution Enabled Residual Networks}
\author{Hao Zhang, \IEEEmembership{Student Member, IEEE}, 
Lu Yuan, 
Guangyu Wu, \\
Fuhui Zhou, \IEEEmembership{Senior Member, IEEE},
and Qihui Wu, \IEEEmembership{Senior Member, IEEE}
\thanks{
H. Zhang, L. Yuan, F. Zhou, and Qihui Wu are with College of Electronic and Information Engineering, Nanjing University of Aeronautics and Astronautics, Nanjing 211106 China. They are also with Key Laboratory of Dynamic Cognitive System of Electromagnetic Spectrum Space (Nanjing University of Aeronautics and Astronautics), and with Ministry of Industry and Information Technology, Nanjing, 211106, China (email: haozhangcn@nuaa.edu.cn, yuanlu@nuaa.edu.cn, zhoufuhui@ieee.org, wuquhui2014@sina.com)

Guangyu Wu is with Department of Computer Science and Technology, University of Science and Technology of China, Hefei 230026 China (email: gywu9908@163.com)
}
}

\begin{document}

\maketitle

\begin{abstract}
Automatic modulation classification (AMC) is of crucial importance for realizing wireless intelligence communications. Many deep learning based models especially convolution neural networks (CNNs) have been proposed for AMC. However, the computation cost is  very high, which makes them inappropriate for beyond the fifth generation wireless communication networks that have stringent requirements on the classification accuracy and computing time. In order to tackle those challenges, a novel involution enabled AMC scheme is proposed by using the bottleneck structure of the residual networks. Involution is utilized instead of convolution to enhance the discrimination capability and expressiveness of the model by incorporating a self-attention mechanism. Simulation results demonstrate that our proposed scheme achieves superior classification performance and faster convergence speed comparing with other benchmark schemes. 
\end{abstract}


\section{Introduction}
\lettrine[lines=2]{A}{UTOMATIC} modulation classification (AMC) is a vital technique to identify the modulation formats under noise and interference \cite{zhang2021novel}. AMC has been widely used in military and civilian applications, such as spectrum management, electronic warfare, interference identification \cite{wu2013spatial,wu2014cognitive}. The AMC schemes mainly have two categories, namely, model-driven AMC and data-driven AMC. The model-driven schemes can be mainly include likelihood-based (LB) schemes and feature-based (FB) schemes \cite{zhang2021novel}. The LB schemes can obtain the optimal solution from the Bayes’ sense by calculating the likelihood function under the modulation hypothesis. The FB schemes aim to find better features and have lower complexity with robust performances. 

Recent studies have demonstrated that data-driven AMC can achieve superior classification performances compared to the model-driven schemes by learning effective representations from data \cite{OShea2018,huynh2020mcnet,huang2019automatic}. Machine learning (ML) based schemes such as support vector machine (SVM), K-nearest neighbor (KNN), and logistic regression can recognize the modulation formats by using a large number of data from the received signals. However, ML-based models still rely on the features generated by FB-based schemes. Thus, deep learning (DL) based algorithms \cite{OShea2018,huynh2020mcnet,huang2019automatic} were proposed to extract features automatically from the original data, such as I/Q samples, cyclic spectrum, constellation diagrams. 

Recently, many novel neural networks were designed to extract discriminative representations for AMC in order to improve the classification performance. A long short-term memory (LSTM) based AMC algorithm was proposed in \cite{rajendran2018deep} to identify modulation formats, but its recurrent structure results in high computational complexity. Inspired by the residual learning for image classification, a modified residual network (ResNet) was applied to extract features from the received I/Q symbols for AMC \cite{OShea2018}. However, the classification performance is limit due to the over-fitting problem caused by a large number of network parameters \cite{huang2019automatic}. Besides those networks, other promising techniques were also used for advancing the performance. For example, a graph convolutional network was investigated for AMC algorithm in \cite{liu2020modulation}. These novel neural networks have improved the classification performance significantly. However, the performance is still limit due to the complex environments, and deep networks need a longer time to converge. In the beyond fifth generation (5G wireless communication networks), intelligent communications with high reliability and low latency are the main characteristics. However, the traditional AMC schemes based on convolution cannot satisfy the classification performance and low computing cost requirements of the beyond 5G wireless communication networks. 

In this letter, we propose a residual network (ResNet) based AMC scheme using involution \cite{li2021involution}. The contributions are summarized as follows. Firstly, involution is utilized instead of convolution to enhance the discrimination capability and expressiveness of the model by incorporating a self-attention mechanism. Secondly, a novel network that inherits the advantages of residual learning and involution is designed to learn high-dimensional representations of different modulations from I/Q signals and construct the classifier. Thirdly, simulation results demonstrate the effectiveness of the proposed involution based ResNet for AMC. It is shown that our proposed scheme has a better classification performance and a faster convergence speed. 

The reminder of this paper is organized as follows. Section II presents the problem of AMC. Section III presents our proposed AMC scheme. Simulation results are given in Section IV and Section V concludes this letter.

\section{Problem Statement}
According to the classical modulation classification problem statement \cite{OShea2018,huynh2020mcnet}, the received signal can be given as
\begin{equation}
x(n)=s(n)+\omega(n), n=1,2,\cdots,N,
\end{equation}
where $N$ is the total number of signal symbols, $s(n)$ denotes the $n$th (complex) symbol, and $\omega(n)$ is the additive white Gaussian noise (AWGN) with zero mean and variance $\delta^2_\omega$. 

The real and imaginary parts of the received signal from the In-phase and Quadrature (I/Q) parts are utilized, which can be expressed as a vector, given as
\begin{align}
\mathbf{x}&=\mathbf{I_x}+\mathbf{Q_x}\nonumber
\\
&=\Re(\mathbf{x})+j\Im(\mathbf{x}),
\end{align}
where $\mathbf{I_x}$ and $\mathbf{Q_x}$ denote the real and imaginary parts of the received signal, respectively, and $j=\sqrt{-1}$. $\Re(\cdot)$ and $\Im(\cdot)$ represent the operators of the real and imaginary parts of the signal, respectively. $\mathbf{x}$ can be specifically expressed as
\begin{equation}
\mathbf{x} = \left(
\begin{array}{c}
\Re[x(1), x(2), \ldots, x(N)]\\
\Im[x(1), x(2), \ldots, x(N)]\\
\end{array} \right).
\end{equation}

The average probability of correct classification ($\Pr_{cc}$) is utilized as the performance metric, which is defined as $\Pr_{cc}=\sum^{|\mathcal{S}|}_{s=1}\Pr(\hat{H} = H_s|H_s)\Pr(H_s), H_s \in \mathcal{S}$, where $\mathcal{S}$ denotes the candidate modulation formats. $\Pr(H_s)$ is the prior probability of modulation format $H_s$, which is equal for each format. $\Pr(\hat{H}=H_s|H_s)\Pr(H_s)$ represents the probability that the modulation format is correctly determined  as $H_s$.

\section{Involution Enabled ResNet for AMC}
In this part, we first introduce the standard convolution operation to make the definition of the proposed involution clearly. Then, involution which inverses characteristics of convolution in the spatial and channel domain with low complexity is presented. Finally, a residual network based on involution for AMC is proposed.

Let $\mathbf{X}\in \mathbb{R}^{H\times W\times C_i}$ denote the input feature map, where $H$, $W$, and $C_i$ represent its height, width and input channels, respectively. A series of convolution filters $C_o$ with the fixed kernel size of $K \times K$ are expressed as $\bm{\mathcal{F}}\in\mathbb{R}^{C_o\times C_i\times K\times K}$, where each filter $\bm{\mathcal{F}}_k\in\mathbb{R}^{C_i\times K\times K}, k=1,2,\cdots,C_o$, contains $C_i$ convolution kernels $\bm{\mathcal{F}}_{k,c}\in\mathbb{R}^{K\times K}, c=1,2,\cdots,C_i$. The filter executes multiply-add operations on the input feature map using a sliding window to generate the output feature map $\mathbf{Y}\in\mathbb{R}^{H\times W\times C_o}$, given as
\begin{equation}
\mathbf{Y}_{i,j,k}=\sum^{C_i}_{c=1}\sum_{(u,v)\in\Delta_K}\bm{\mathcal{F}}_{k,c,u+\lfloor K/2\rfloor,v+\lfloor K/2\rfloor}\mathbf{X}_{i+u,j+v,c},
\end{equation}
where $\Delta_K\in\mathbb{Z}^2$ denotes the set of offsets in the neighborhood considering convolution conducted on the center pixel, given as
\begin{equation}
\Delta_K=[-\lfloor K/2\rfloor,\cdots,\lfloor K/2\rfloor]\times[-\lfloor K/2\rfloor,\cdots,\lfloor K/2\rfloor].
\end{equation}

It is well known that there is inter-channel redundancy inside convolution filters, which results in the flexibility problem in convolution operation \cite{li2021involution}. Compared to the standard convolution, the \textbf{involution} kernel $\bm{\mathcal{H}}\in\mathbb{R}^{H\times W\times K\times K\times G}$ is designed to realize transforms with inverse characteristics in the spatial and channel domain. Specifically, an involution kernel $\bm{\mathcal{H}}_{i,j,\cdot,\cdot,g}\in\mathbb{R}^{K\times K}, g=1,2,\cdots,G$, is specially adapted for the pixel $\mathbf{X}_{i,j}\in\mathcal{R}^C$ located at the corresponding coordinate $(i, j)$, but shared over the channels. $G$ denotes the number of groups and each group shares the same involution kernel. The output feature map of involution is derived by executing multiply-add operations on the input with the involution kernels, given as
\begin{equation}
\mathbf{Y}_{i,j,k}=\sum_{(u,v)\in}\bm{\mathcal{H}}_{i,j,u+\lfloor K/2\rfloor,v+\lfloor K/2\rfloor,\lceil kG/C\rceil}\mathbf{X}_{i+u,j+v,k}.
\label{eq6}
\end{equation}

\begin{figure}[t]
	\includegraphics[width=0.95\linewidth]{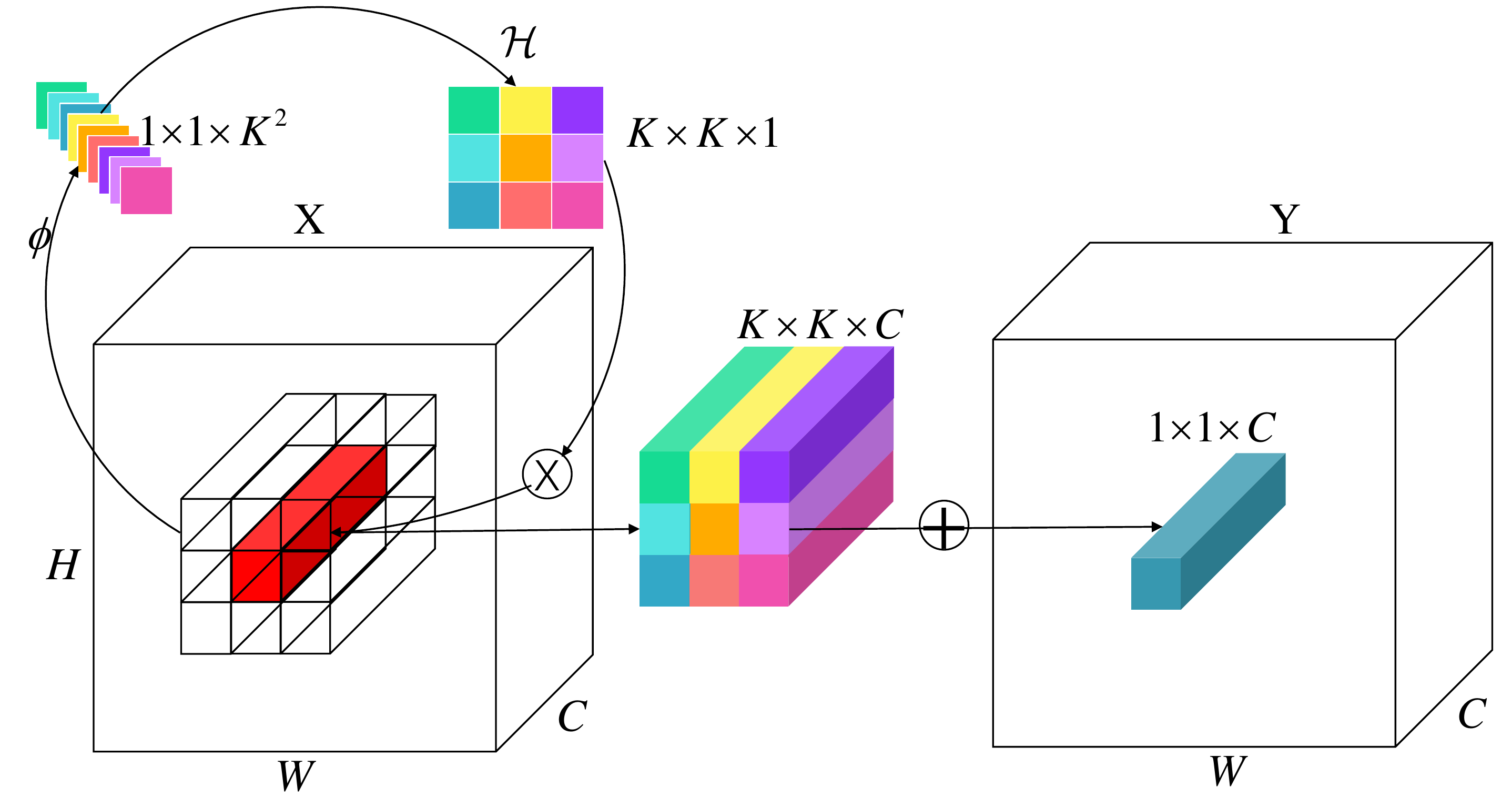}
	\caption{Schematic illustration of the proposed involution. 
	The involution kernel $\mathcal{H}_{i;j}\in\mathbb{R}^{K\times K\times 1}$ ($G = 1$ for ease of demonstration) is produced from the function $\phi$ based on a single pixel at $(i, j)$, followed by a channel-to-space rearrangement. 
	The multiply-add operation of involution consists of two steps, where $N$ indicates multiplication broadcast across $C$ channels and $L$ represents summation aggregated within the $K \times K$ spatial neighborhood. }
	\label{fig:involution}
\end{figure}

Different from convolution kernels, the shape of involution kernels $\bm{\mathcal{H}}$ depends on the input feature map $\mathbf{X}$. The output kernels are aligned to the input by generating the involution kernels based on the original input tensor. Thus, the kernel generation function $\phi$ and mapping function at each location ($i, j$) can be expressed as 
\begin{equation}
\bm{\mathcal{H}}_{i,j}=\phi(\mathbf{X}_{\Psi_{i,j}}),
\end{equation}
where $\Psi_{i,j}$ indexes the set of pixels $\bm{\mathcal{H}}_{i,j}$ is conditioned on. The kernel generation function $\phi:\mathbb{R}^C\to\mathbb{R}^{K\times K\times G}$ with $\Psi_{i,j}=\{(i,j)\}$ is given as
\begin{equation}
\bm{\mathcal{H}}_{i,j}=\phi(\mathbf{X}_{i,j})=\mathbf{W}_1\sigma(\mathbf{W}_0\mathbf{X}_{i,j}),
\label{eq8}
\end{equation}
where $\mathbf{W}_0\in\mathbb{R}^{\frac{C}{r}\times C}$ and $\mathbf{W}_1\in\mathbb{R}^{(K\times K\times G) \times\frac{C}{r}}$ denote two linear transformations that collectively constitute a bottleneck structure. The channel reduction operation under a ratio $r$ is used for efficient processing, and $\sigma$ represents Batch Normalization and non-linear activation functions between two linear projections. The pseudo code of Algorithm \ref{alg:involution} shows the computation flow of involution, which is visualized in Fig. \ref{fig:involution}. 

\begin{algorithm} 
\caption{Pseudo code of involution.} 
\label{alg:involution} 
\begin{algorithmic}[1] 
\REQUIRE Batch size $B$, height $H$, width $W$, channel $C$, group number $G$, kernel size $K$, stride $s$, and reduction ratio $r$;
\ENSURE The involution kernel $out$.
\STATE \textbf{Initialize the network operations};
\STATE Define the operation $o$ as average pooling with kernel $s$ if $s>1$, otherwise $o$ is the identity mapping.
\STATE Define the operation $reduce$ as convolution with kernel of $C\times C//r\times 1$.
\STATE Define the operation $span$ as convolution with kernel of $C//r\times K*K*G\times 1$.
\STATE Define the operation $unfold$ as unfold with kernel of $K\times dilation\times padding\times s$.
\STATE \textbf{Forward pass};
\STATE Calculate $x\_u$ using $x\_u = unfold(x)$, and reshape it into $(B, G, C//G, K*K, H, W)$;
\STATE Generate the involution kernel using eq. \ref{eq8} as $kernel = span(reduce(o(x)))$, and reshape it into $(B, G, K*K, H, W)$;
\STATE Execute Multiply-Add operation according to eq. \ref{eq6} as $out = mul(kernel, x\_u).sum(dim=3)$, and reshape it into $(B, C, H, W)$.
\end{algorithmic}
\end{algorithm}

The feature generation process in eq. \ref{eq6} can be considered a generalized version of self-attention \cite{vaswani2017attention}. The self-attention pools \emph{values} $\bm{V}$ depending on the affinities obtained by computing similarity between the \emph{query} $\bm{Q}$ and \emph{key} $\bm{K}$, are formulated as
\begin{equation}
\bm{Y}_{i,j,k}=\sum_{(p,q)\in\Omega}(\bm{QK}^\top)_{i,j,p,q,\lceil kH/C\rceil}\bm{V}_{p,q,k},
\end{equation}
where $\bm{Q}$, $\bm{K}$, and $\bm{V}$ are linearly transformed from the input $\bm{X}$, and $H$ is the number of heads in multi-head self-attention \cite{vaswani2017attention}. The similarity is related to that both operators collect pixels in the neighborhood $\delta$ or a less bounded range $\Omega$ through a weighted sum. On one hand, the computation of involution can be viewed as a spatial attentive aggregation. On the other hand, the attention map (also called affinity or similarity matrix) $\bm{QK}^\top$ in the self-attention mechanism can be considered as a kind of involution kernel $\bm{\mathcal{H}}$.

\begin{figure}[t]
	\includegraphics[width=0.95\linewidth]{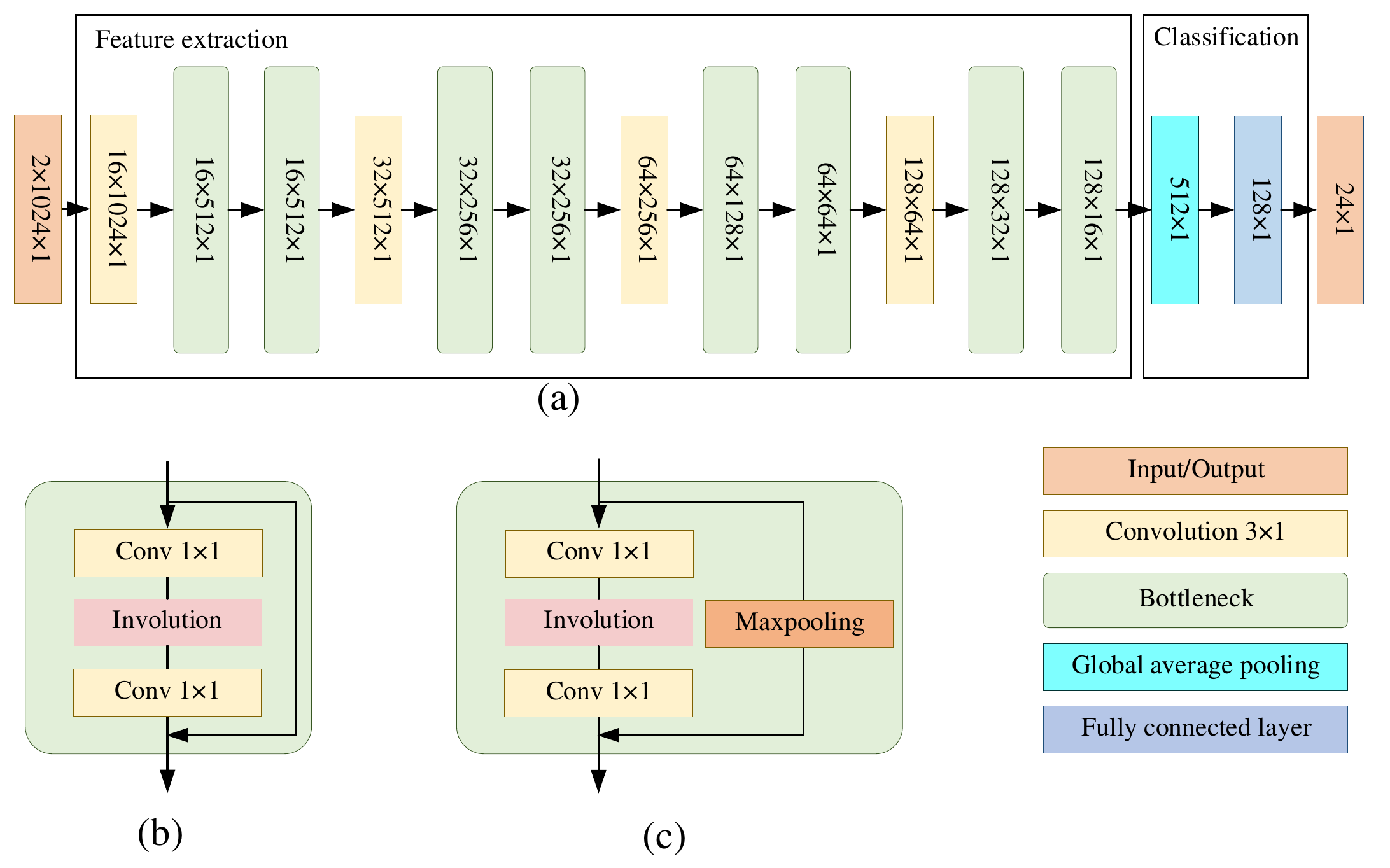}
	\caption{Structure of the proposed Invo-ResNet, (a) network structure, (b) bottleneck, (c) bottleneck with downsampling using maxpooling.}
	\label{fig:model}
\end{figure}

To build an entire network with involution for AMC, we design a novel lightweight network named Invo-ResNet by stacking bottleneck blocks and using involution kernels since the elegant architecture of ResNet makes it successful for many applications \cite{qi2020automatic}. As shown in Fig. \ref{fig:model}, the proposed Invo-ResNet consists of two modules, namely, the feature extraction module and classification module. The feature extraction module aims at extracting the underlying high-level representations from the input signals. Then, these representations are transformed into fixed-dimensional feature vectors by the global average pooling (GAP) layer. Modulation classification is subsequently conducted using these vectors in the fully connected (FC) layer of the classification module.

To balance the tradeoff between accuracy and efficiency, the feature extraction module consists of one convolutional layer and two bottlenecks. A convolutional layer with a kernel size of $3\times 1$ is used to extract low-level information and execute the feature aggregation. Then, the bottleneck is utilized to learn high-level features by using involution with fewer parameters compared to convolution. As shown in Fig. \ref{fig:model} (b), the bottleneck is constructed by replacing the $3\times 3$ convolutional layer in the original ResNet \cite{OShea2018} with involution and retaining all the $1\times 1$ convolution for channel projection and fusion \cite{li2021involution}. Instead of using a constant channel number as in \cite{OShea2018}, a pyramid architecture with increasing channel numbers is utilized in the Invo-ResNet. Therefore, maxpooling is adopted to decrease the shape of the feature map in order to reduce the computation cost, as shown in Fig. \ref{fig:model}(c). 

For the last part, a GAP layer and an FC layer are served as the classification module. GAP layer aggregates information from the feature extraction module and enables the network to take samples with an arbitrary length. \emph{Softmax} is used as the activation function in the last FC layer to normalize the output of each neuron, indicating the probability that the target signal belongs to the corresponding modulation format. Moreover, we adopt rectified linear units (ReLU) as the activation function in the convolutional layers to introduce nonlinearity and sparsity.

\section{Simulation}

The pubic RadioML 2018.01A that contains $24$ kinds of modulations under an SNR range from $-20$dB to $30$dB with a step of $2$dB is adopted. There are over $2$ million samples with 1024 points. $80\%$ of these data are utilized for training and the rest $20\%$ are used for testing. The models are trained by using SGD with an initial learning rate of $0.01$, a weight decay of $5 \times 10^{-4}$, and a momentum of $0.9$ for $50$ epochs. 

\begin{figure*}[h]
\centering
\subfigure[]{
\begin{minipage}[h]{0.45\linewidth}
\centering
\includegraphics[width=0.85\linewidth]{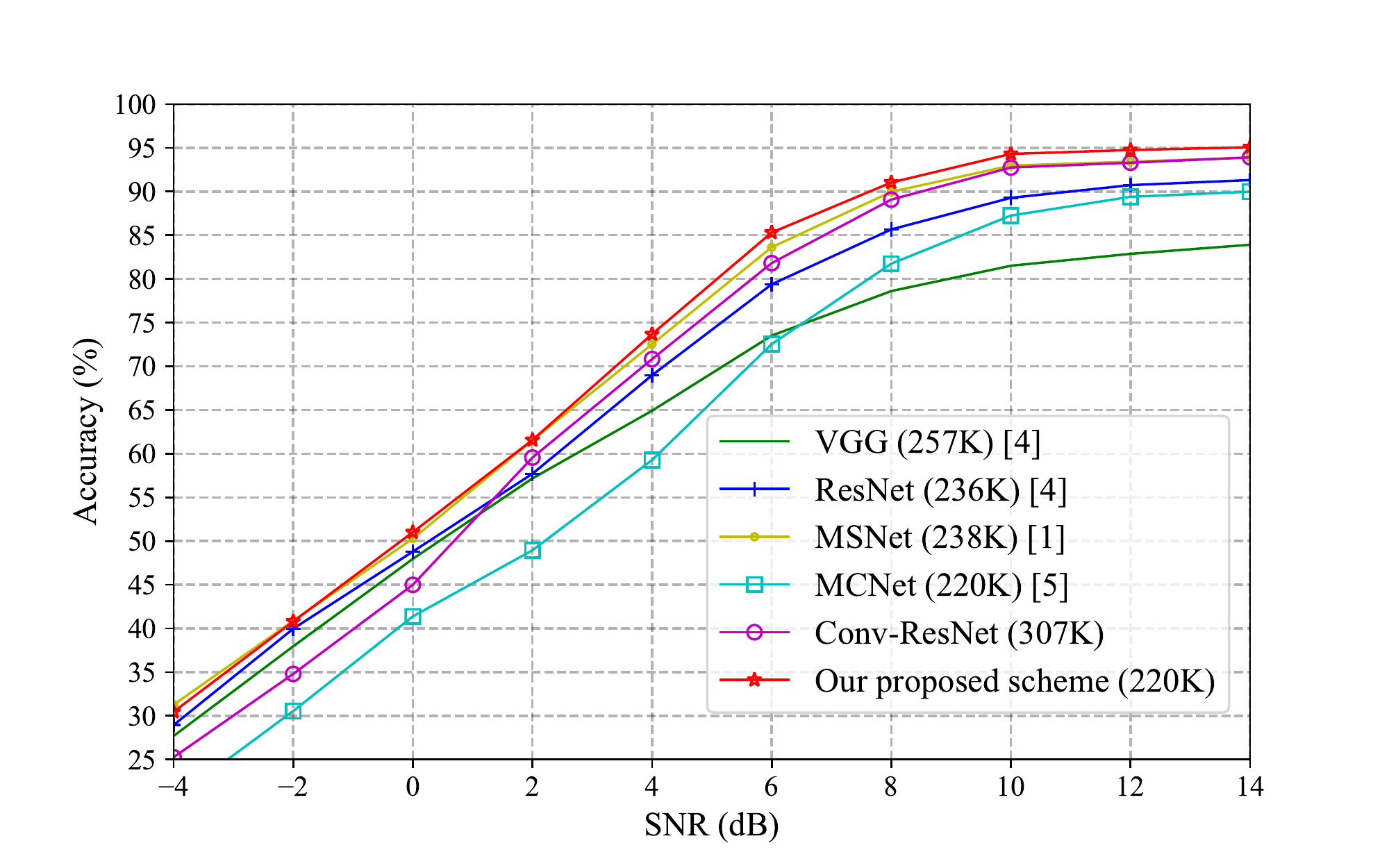}
\label{fig:comparison}
\end{minipage}
}
\subfigure[]{
\begin{minipage}[h]{0.45\linewidth}
\centering
\includegraphics[width=0.85\linewidth]{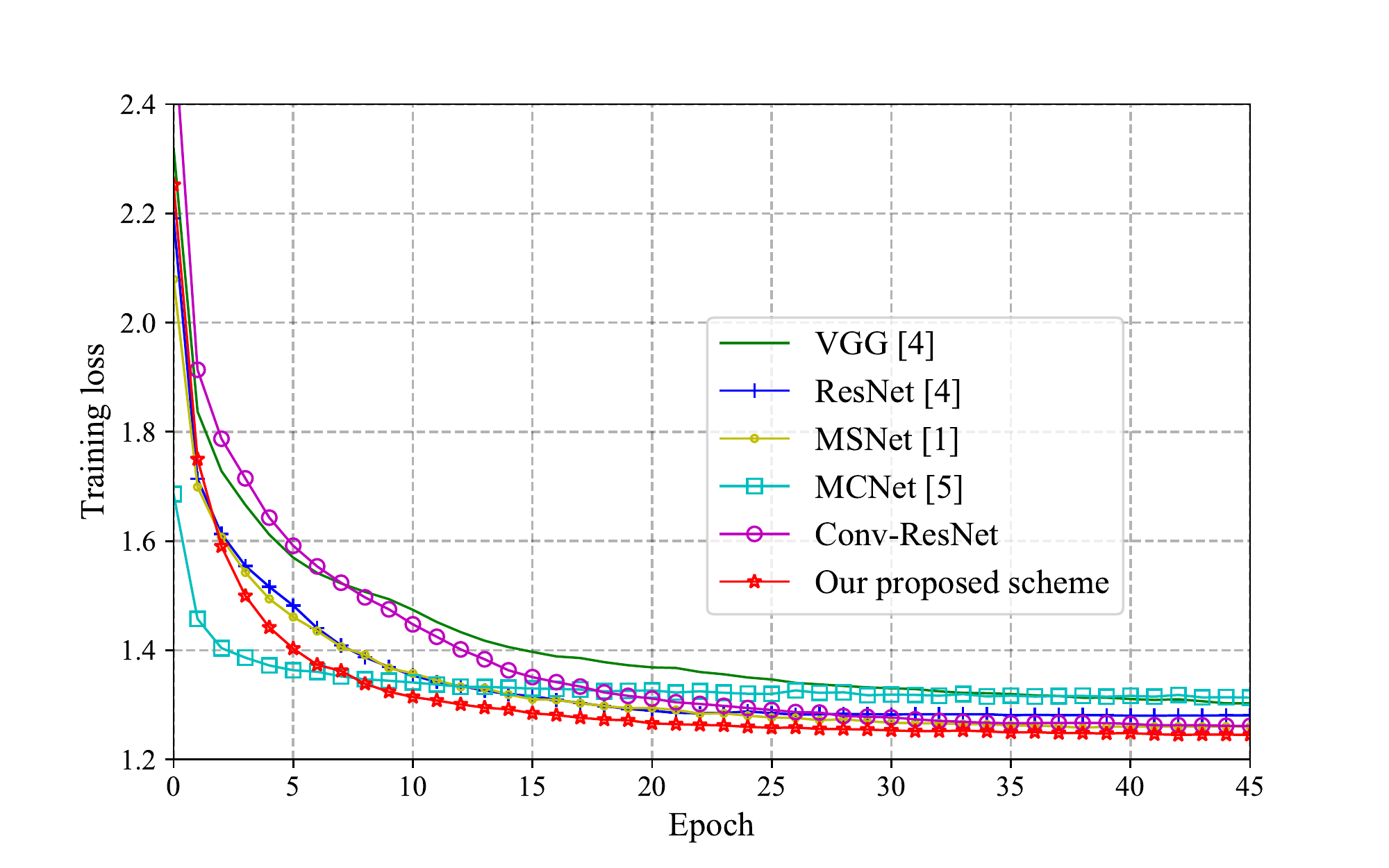}
\label{fig:loss}
\end{minipage}
}
\caption{Comparison of (a) the classification performance and (b) the training loss among VGG, ResNet, MSNet, {\color{blue} MCNet, our proposed scheme with convolution layers and our proposed scheme with involution layers}.
}\label{fig:result}
\end{figure*}

Fig. \ref{fig:comparison} shows the classification performance comparison of our proposed proposed Invo-ResNet with those achieved by three representative DL-based models for AMC including VGG \cite{OShea2018}, ResNet \cite{OShea2018} and MCNet \cite{huynh2020mcnet} on RadioML 2018.01A dataset. It is evident that the proposed Invo-ResNet is superior to other traditional models, and it can provide $2\%$ gains over our previous work MSNet, $2$ dB gains over ResNet, $3$ dB gains over MCNet and $4$ dB gains over VGG. Moreover, Invo-ResNet can reach over $90\%$ accuracy when the SNR is larger than $6$dB, and it can achieve about $95\%$ accuracy at $10$dB, while the best performance of ResNet is $91.47\%$ at $10$dB. To further show the superiority of the proposed Invo-ResNet, Fig. \ref{fig:loss} illustrates the training loss during the training process of the compared models. The proposed Invo-ResNet can achieve a lower loss in the training set compared to the other models and obtains a faster convergence speed than other models. To show the effectiveness of the involution for AMC, a fully convolutional network with the same structure is trained and tested under the same dataset. As shown in Fig. \ref{fig:result}, compared to our proposed scheme with involution, the convolutional counterpart (Conv-ResNet) only reaches a best performance of $94.5\%$ at high SNR condition, which is about $1\%$ lower than involution based model, and it performs worse than our previous work MSNet. For the analysis of the computational complexity, we calculate the network parameters of the compared models, as shown in Fig. \ref{fig:comparison}. Our proposed scheme achieves a better performance with less parameters than these conventional schemes. Moreover, compared to the convolution based scheme under the same structure, our proposed scheme with involution has about $40\%$ reduced parameters, which can be implemented in resource limited devices.

\begin{figure*}[h]
\centering
\subfigure[]{
\begin{minipage}[t]{0.3\linewidth}
\centering
\includegraphics[width=0.65\linewidth]{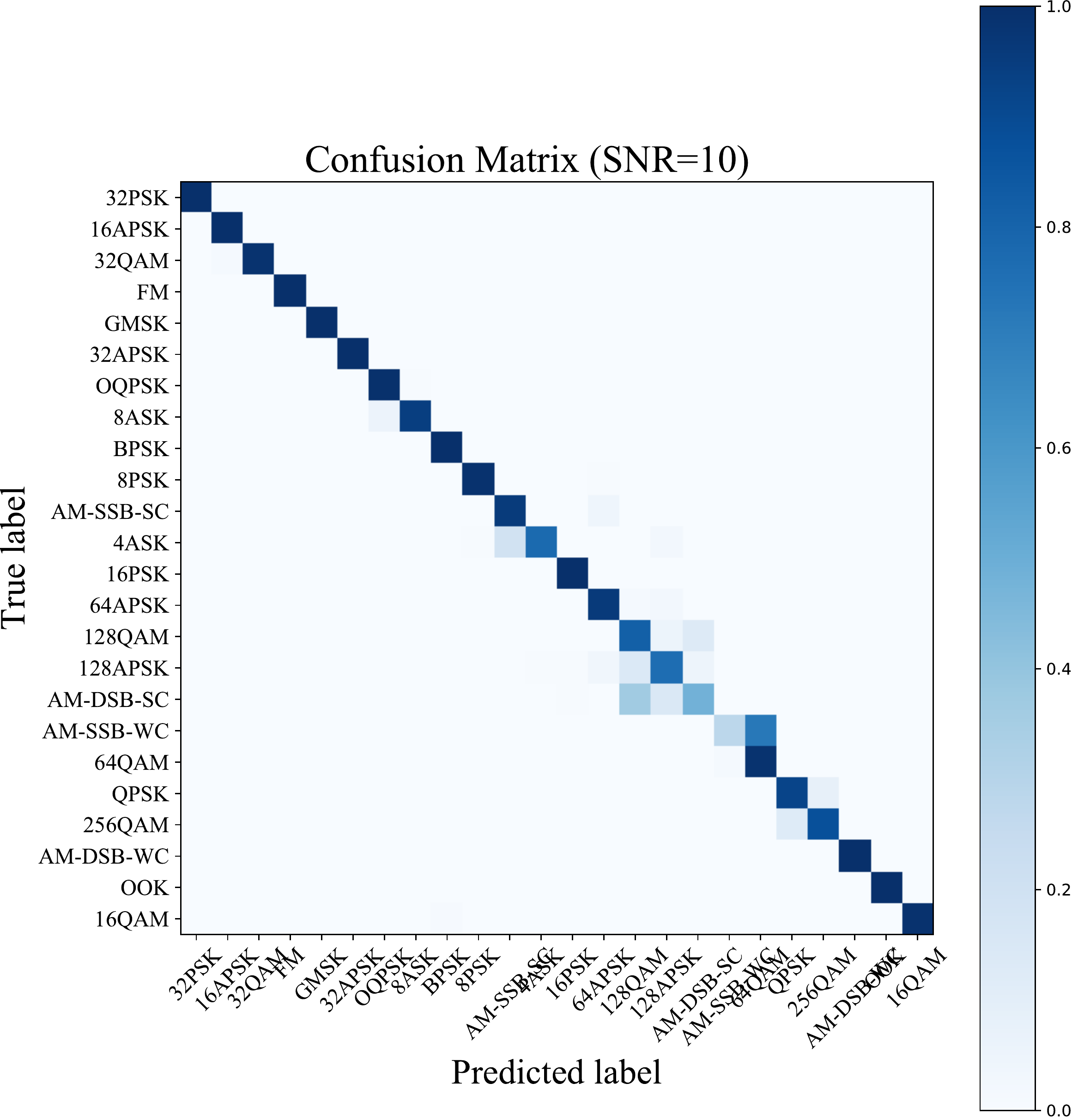}
\label{fig:resnet}
\end{minipage}
}
\subfigure[]{
\begin{minipage}[t]{0.3\linewidth}
\centering
\includegraphics[width=0.65\linewidth]{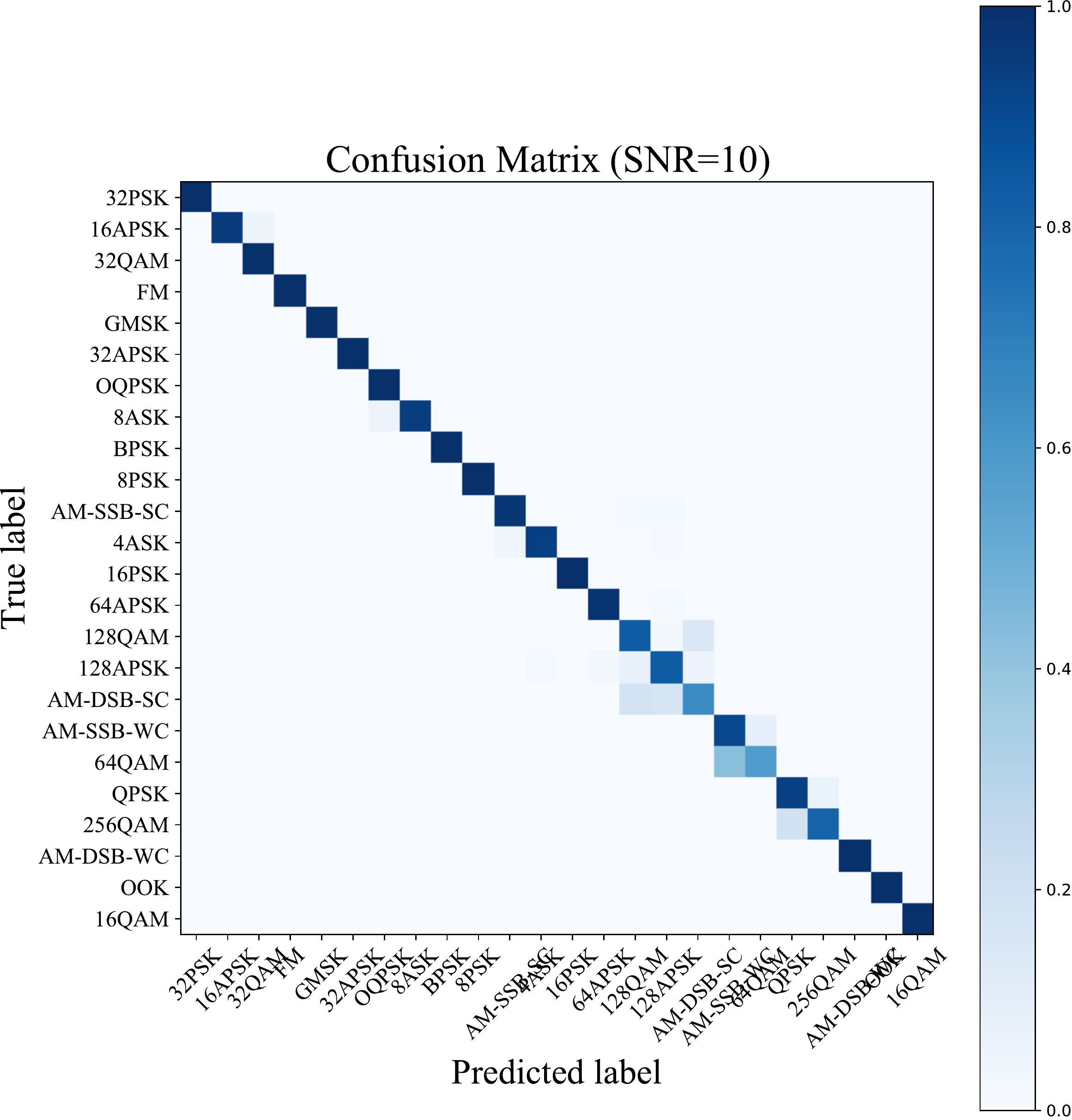}
\label{fig:msnet}
\end{minipage}
}
\subfigure[]{
\begin{minipage}[t]{0.3\linewidth}
\centering
\includegraphics[width=0.65\linewidth]{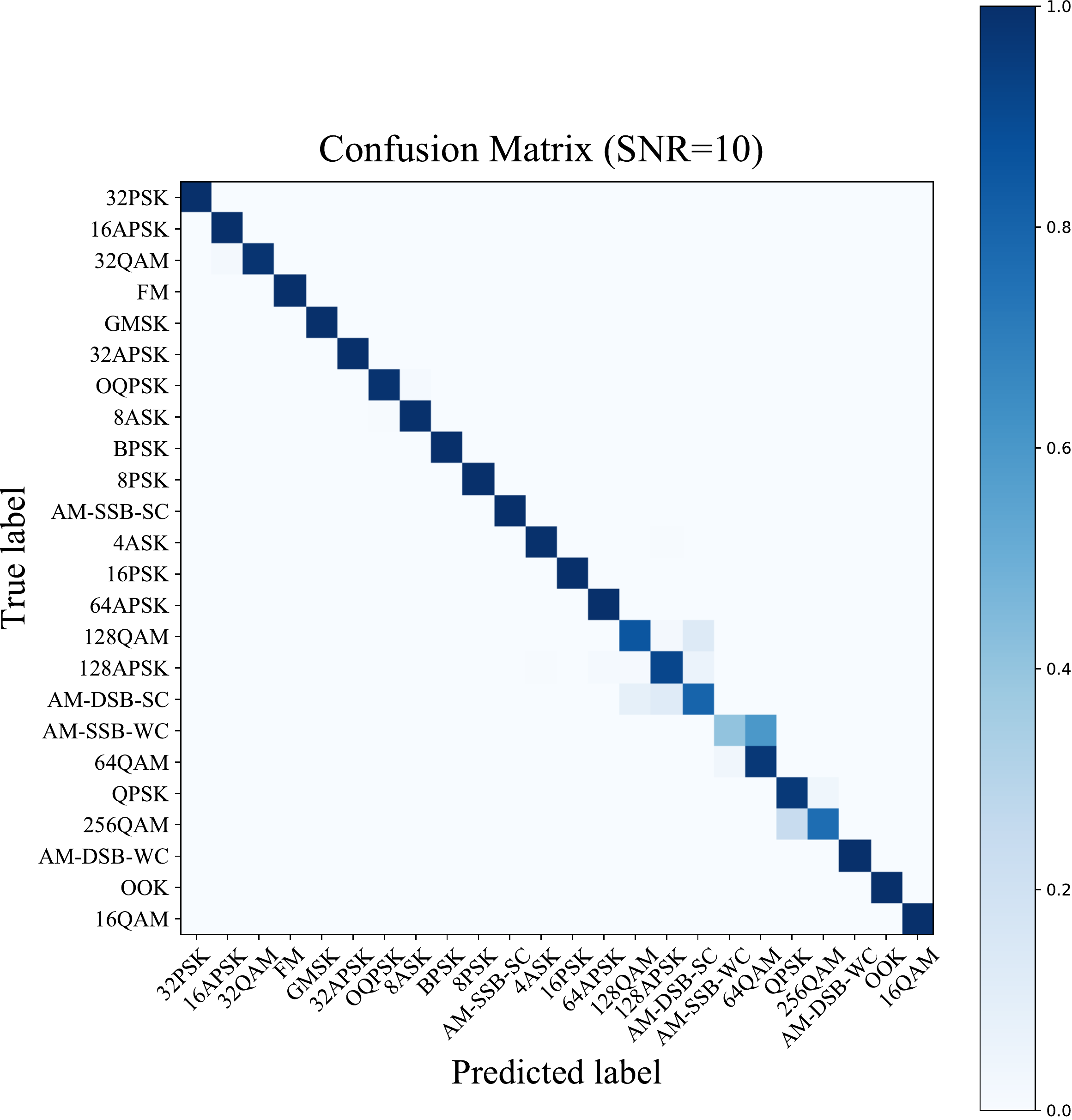}
\label{fig:invo}
\end{minipage}
}
\caption{Comparison of confusion matrix at $10$dB, (a) ResNet \cite{OShea2018}, (b) MSNet \cite{zhang2021novel}, and (c) Our proposed scheme. 
}\label{fig:confusion}
\end{figure*}

To further demonstrate the effectiveness of the proposed Invo-ResNet, the comparison of the confusion matrices at $10$dB of ResNet \cite{OShea2018}, MSNet \cite{zhang2021novel} with those of our proposed scheme are shown in Fig. \ref{fig:confusion}. It is seen that the proposed Invo-ResNet has less confusion compared to other traditional models. Specifically, only AM-DSB-WC receives the worst performance in our proposed scheme. On the contrary, two modulation formats perform worse in ResNet and MSNet, which are AM-DSB-SC and $64$QAM, and AM-DSB-SC and AM-SSB-WC as shown in Fig. \ref{fig:resnet} and Fig. \ref{fig:msnet}, respectively.

\section{Conclusion}
A novel AMC scheme was proposed by designing a novel network using involution. In order to improve the classification accuracy and decrease the computation cost of convolution based AMC schemes, involution was utilized to enhance the discrimination capability and expressiveness of the model and reduce the training time by incorporating a self-attention mechanism. Simulation results demonstrated the superiority of our proposed scheme in terms of classification accuracy and the training time. In this case, our proposed scheme is more appropriate in beyond 5G wireless communication networks

\bibliographystyle{IEEEtran}

\end{document}